\documentclass[twocolumn,showpacs,amsmath,prl,amssymb]{revtex4}

\usepackage{bm}

\begin{document}

\title{Mesoscopic charging effects in the $\sigma$-model for granular
  metals}

\author{A.~V.~Andreev} \affiliation{Department of Physics, University of
  Colorado, CB 390, Boulder, Colorado 80390} \affiliation{Bell
  Laboratories, Lucent Technologies, 600 Mountain Avenue, Murray Hill, New
  Jersey 07974}

\author{I.~S.~Beloborodov}
\altaffiliation[Present address:]{Argonne National Laboratory, Argonne,
  Illinois 60439. }
\affiliation{Department of Physics, University of Colorado, CB 390,
  Boulder, Colorado 80390} \affiliation{Bell Laboratories, Lucent
  Technologies, 600 Mountain Avenue, Murray Hill, New Jersey 07974}

\date{\today}

\begin{abstract}
  We derive the $\sigma$-model for granular metals. It is valid for any
  relation between the temperature, $T$, and the grain mean level spacing.
  The continuum limit of the $\sigma$-model describes non-perturbative
  charging effects in homogeneous disordered metals.  The $\sigma$-model
  action contains a novel term which is crucial for studying Coulomb
  blockade effects.  Topological properties of the $\sigma$-model encoding
  charge discreteness are studied.  For $T$ below the escape rate from the
  grain, $\Gamma$, Coulomb blockade effects are described by solitons of
  the $Q$-matrix which are delocalized in real space. At $T> \Gamma $ the
  solitons transform into phase instantons localized at a single tunneling
  contact.
\end{abstract}


\maketitle

Physics of electron-electron interactions in mesoscopic systems
has been the subject of intensive experimental and theoretical
studies~\cite{Altshuler85}.  Recently, much attention was
attracted to the study of granular
metals~\cite{Efetov,experiment,Efetov02,Lopatin03,Lerner01,we2002,Kamenev03}.
They can be viewed as extended networks of metallic grains
connected by tunneling contacts with dimensionless conductance
$g_T$.  The Coulomb interaction in such systems may be described
by capacitive coupling between the grains,
\begin{equation}
  \label{eq:cint1}
  \hat{H}_C= \frac{e^2}{2}\sum \limits_{{\bf l},{\bf l}'}
  \left(\hat{N}_{\bf l} - q_{\bf l}\right)C_{{\bf l}{\bf l}'}^{-1}
  \left(\hat{N}_{{\bf l}'} - q_{{\bf l}'}\right).
\end{equation}
Here $e$ is the electron charge, ${\bf l}$ and $ {\bf l}'$ label the
grains, $C_{{\bf l}{\bf l}'}$ is the capacitance matrix, $\hat{N}_{\bf l}$
is the electron number operator in the ${\bf l}$-th grain, and $q_{\bf l}$
are linear in external electrostatic potentials, such as gate voltages.

At temperatures below the charging energy, $E_C \sim e^2/C$, the
fluctuations of the number of electrons in the grain are
suppressed and electron transport is inhibited due to Coulomb
blockade effects.  The latter are typically
treated~\cite{Shoenreview,Efetov02,Kamenev03} using the
dissipative action of Ref.~\cite{AES} in terms of the phase
$\phi(\tau)$ conjugate to the grain charge.  Coulomb blockade
effects are non-perturbative in the tunneling conductance $g_T$
and are described by the phase instantons~\cite{Korshunov}. The
approach of Ref.~\cite{AES} neglects elastic propagation of
electrons across the grain (elastic cotunneling) and is valid for
zero mean level spacing, $\delta \to
0$~\cite{Averin90,Matveev95,Aleiner97}. At temperatures smaller
than the escape rate from the grain, $\Gamma\sim g_T \delta$,
elastic propagation of electrons becomes important. In this regime
the dissipative action of Ref.~\cite{AES} should be replaced by
the $\sigma$-model in which electron propagation across grains is
described by the $Q$-matrix.

For $g_T \gg 1$ the electron transport at $T \ll \Gamma$ is
diffusive~\cite{Efetov}. Granular metals in this regime may be described
by the continuum limit of the $\sigma$-model.  It is expected that the
continuum limit of the $\sigma$-model is independent of the granular
nature of the metal.  Therefore, studying
charging effects in this regime will shed light on the role of charge
discreteness effects in homogeneous disordered metals which are expected to become
important close to the metal-insulator transition.  In the metallic regime
Coulomb blockade effects are non-perturbative in $1/g_T$ and in the
$\sigma$-model treatment should be described by $Q$-matrix solitons which
replace the phase instantons at $T < \Gamma$.  An important step in this
direction was made in Ref.~\cite{Kamenev00}. Since the consideration of
Ref.~\cite{Kamenev00} assumes zero mean level spacing in the grain and
spatial separation between the Coulomb interaction and $Q$-matrix degrees
of freedom it is not clear how to generalize it to the continuum limit.

Here we derive the $\sigma$-model for granular metals,
Eq.~(\ref{sigmamodel_ab}).  It enables one to study
non-perturbative charging effects at $T < \Gamma$. In contrast to
Refs.~\cite{Efetov,Lerner01} its applicability is not limited by
the condition $T \gg \delta$.   The continuum limit of the
$\sigma$-model, Eqs.~(\ref{eq:F_d_cont}) and (\ref{eq:F_C_cont}),
describes  charging effects in homogeneous disordered metals.
Equations (\ref{sigmamodel_ab}), (\ref{eq:F_d_cont}) and
(\ref{eq:F_C_cont})
 represent our main results.

We consider a $d$-dimensional array of metallic grains coupled by
tunneling contacts with dimensionless conductance $g_T = \frac{2\pi
  \hbar}{e^2 R}$, with $R$ being the contact resistance. We assume that
the Thouless energy of the grain, $E_T$, exceeds both $E_C$ and the
elastic escape rate, $\Gamma$. We make no assumptions about the relation
between $T$ and $\delta$, where $\delta$ is the mean level spacing of the
grain per spin projection.  For simplicity of presentation we restrict
ourselves to the case of normal metals with broken time reversal symmetry.
The generalization to other symmetry classes and superconducting case will
be considered elsewhere.
We consider electrons moving in the presence of a random
potential, $V_{\bf l}({\mathbf r})$, with a variance $\langle
V_{\bf l}({\mathbf r})V_{{\bf l}'}({\mathbf r'})\rangle=
\frac{1}{2\pi \nu \tau} \delta({\mathbf r}- {\mathbf
r'})\delta_{{\bf
    l}{\bf l}'}$, where $\nu$ is the density of states and $\tau$ is the
mean free time.  Electron tunneling between adjacent grains is described
by the tunneling Hamiltonian
\begin{equation}
\label{hamiltun}
\hat{H}_t =  t \sum\limits_{{\bf l}{\bf l}', p k}
c_{{\bf l}{ p}}^{\dagger}c^{}_{{\bf l}'{  k}} + h.c.,
\end{equation}
where $c^{\dagger}_{{\bf l}p}$ is the electron creation operator.
Finally, the Coulomb interaction is described by Eq.~(\ref{eq:cint1}).

Thermodynamic and linear response transport properties of a granular metal
can be obtained from the average replicated partition function, $ \langle
Z^\alpha \rangle=\left\langle
  \exp\left(-\alpha\frac{\hat{H}}{T}\right)\right\rangle$, where $\alpha$
is the number of replicas and the averaging is performed over the random
impurity potential. We write $\langle Z^\alpha \rangle$ as an imaginary
time functional integral over the fermions.  Averaging over disorder
induces a quartic interaction of fermions which is decoupled using a
Hermitian matrix field $\tilde{Q}$~\cite{Finkelstein83,Efetov97}. The
Coulomb interaction is decoupled by the scalar field,
$V_{j\textbf{l}}(\tau)$, in each replica.  Integrating out the fermions we
obtain,
\begin{subequations}
\label{eq:Z_abc}
\begin{eqnarray}
\label{eq:Z}
\langle Z^{\alpha}\rangle &=&  \int d[\tilde{Q},V]
e^{ -  \frac{\pi \nu}{4\tau}{\rm Tr} \tilde{Q}^2 -
{\cal F}_d -  {\cal F}_C }, \\
\label{eq:F_C1}
{\cal F}_C&=&  \sum\limits_{j\bf{l}\bf{l}'}
 \int\limits_0^{\beta} d\tau
  V_{j\bf{l}}\left[
\frac{C_{\bf{l}\bf{l}'}}{2e^2}
V_{j\bf{l}'}  +  i  \delta_{\bf{l}\bf{l}'} q_{\bf{l}} \right]  ,\\
\label{eq:F}
{\cal F}_d &=& - {\rm Tr} \ln \left( i\hat{\epsilon} +
i V  -\hat{t}  -\hat{H}_0+\frac{i\tilde{Q}}
  {2\tau}  \right)  .
\end{eqnarray}
\end{subequations}
Here $\hat \varepsilon = i \partial_\tau$, and $\hat{H}_0=-\mu -\frac{\nabla^2}{2m}$,
with $\mu$ being the chemical
potential, is the single particle Hamiltonian in the absence of tunneling
and disorder.  The operator $\hat{t}$ describes tunneling between grains,
and we omitted the argument of the scalar potential $V_{j\bf{l}}(\tau)$.
The trace in Eq.~(\ref{eq:F}) is taken over the
spin, replica, Matsubara and grain indices.

It is convenient to introduce the phases, $\chi_{j{\bf l}}(\tau)$, by
singling out the static part, $V^0_{j{\bf l}}$, of the auxiliary field,
\begin{equation}
V_{j{\bf l}}(\tau) = V^0_{j{\bf l}}+ \dot\chi_{j{\bf l}}(\tau),
\label{eq:chi_def}
\end{equation}
where $\chi_{j{\bf l}}(0)=\chi_{j{\bf l}}(\beta)$.  The static part,
$V^0_{j{\bf l}}$, plays the role of fluctuating chemical potential and is
responsible for charge neutrality. The time-dependent part of $V_{j{\bf
    l}}(\tau)$ can be removed from Eq.~(\ref{eq:F}) by the gauge
transformation,
\begin{subequations}
\label{eq:gauge_ab}
\begin{eqnarray}
\label{eq:gauge_a}
\tilde{Q}(\tau,\tau') &=& \exp[ i \chi(\tau)]
Q(\tau,\tau')\exp[-i \chi(\tau ')], \\
\label{eq:gauge_b}
t& \to& t \exp[-i\hat{\chi}_{j{\bf ll}'}(\tau) ],
\end{eqnarray}
\end{subequations}
where $\hat{\chi}_{\bf{ll}'}$ denotes a diagonal matrix of phase differences,
$\hat{\chi}_{\bf{ll}'}(\tau) = \delta_{ij}[\chi_{j\bf{ l}}(\tau)
-\chi_{j\bf{l}'}(\tau)] $.

Following the standard procedure~\cite{Finkelstein83,Efetov97} we look for
diagonal coordinate-independent saddle points of the action in
Eq.~(\ref{eq:Z}). The tunneling Hamiltonian in Eq.~(\ref{eq:F}) leads to a
finite escape rate from the grain, $\Gamma \sim t^2/\delta$. For $\Gamma
\ll 1/\tau$ this term may be ignored while looking for the saddle point,
\begin{equation}
\label{spa}
Q_{j{\bf l}}= \frac{i}{\pi \nu}\left. \left( i\hat{\epsilon}
 + i V^0_{j{\bf l}} -\hat{H}_0
 + \frac{i Q_{j{\bf l}}}{2\tau}  \right)^{-1}\right|_{{\bf r}={\bf r}'}  .
\end{equation}
We denote by $\Lambda$ the solution of Eq.~(\ref{spa}) which for
$\epsilon_n \ll \mu$ has the form, $Q_{j{\bf
    l}}(\epsilon_n)=\Lambda(\epsilon_n) = \textrm{sign}(\epsilon_n)$.  It
has the meaning of the average equal point Green's function of an isolated
grain with the chemical potential $\mu +i V^0_{j{\bf l}}$ .

At high energies, $|\epsilon_n| \gg 1/\tau$ the fluctuations of the
$Q$-matrix about the saddle point are strongly suppressed, and it can be
replaced by $Q=\Lambda$. Using this we separate the high energy part of
the action, Eq.~(\ref{eq:F}), and rewrite the effective low-energy action
in the form,
\begin{eqnarray}
{\cal F}_d &=& {\cal F}_0[V]-\frac{g_{T}}{8}
\sum\limits_{\textbf{l}\textbf{l}'}
{\rm Tr} [ e^{i\hat{\chi}_{\bf{ll}'}}
\Lambda_{\bf{l}}e^{-i\hat{\chi}_{\bf{ll}'}}\Lambda_{\bf{l}'}]   \nonumber \\
&&-{\rm Tr}' \ln \left( \frac{i\hat{\epsilon} + i V^0
  - \hat{t}-\hat{H}_0+\frac{i Q}
  {2\tau} }{i\hat{\epsilon} + i V^0
  - \hat{t}-\hat{H}_0+\frac{i \Lambda}
  {2\tau}} \right) .
  \label{eq:F2}
\end{eqnarray}
To arrive at Eq.~(\ref{eq:F2}) we expanded the right hand side of
Eq.~(\ref{eq:F}) to second order in $t$. Here ${\rm Tr}'$ denotes
the trace restricted to the low-energy sector, $|\epsilon_n| \leq
1/\tau$, and ${\cal F}_0[V]$ denotes the action ${\cal F}_d$,
Eq.~(\ref{eq:F}), in the absence of tunneling, $t=0$. We may
expand ${\cal F}_0[V]$ to second order in $V^0$. Taking into
account that the electron charge density is cancelled by that the
ionic background we obtain, ${\cal F}_0[V]= {\cal F}_0[0] +\sum_{j
{\bf l}}(V^0_{j{\bf l}})^2/(T \delta) $.

The action, Eq.~(\ref{eq:F2}), is minimized at the saddle point given by
the solution of Eq.~(\ref{spa}). The most general diagonal solution is
given by $Q_0(\epsilon_n)=\pm 1$. The massive fluctuations about the
saddle point can be integrated out in the Gaussian approximation. The
remaining integral over the massless modes constitutes the $\sigma$-model.
For energies $\epsilon \ll E_T$ the massless modes are parameterized by
$Q$-matrices which are coordinate independent within each grain (zero-mode
approximation). They can be expressed as
\begin{equation}
 Q_{\textbf{l}}=\hat{U}_{\textbf{l}}^\dagger Q_0 \hat{U}_{\textbf{l}},
\label{eq:parametrization}
\end{equation}
where $Q_0$ is a diagonal matrix with diagonal elements equal to
$\pm 1$, and $\hat{U}_{\textbf{l}}$ are unitary matrices which
differ from unity only for $|\epsilon|, |\epsilon'| < E_T$. Thus
the manifold of massless modes breaks up into disjoint manifolds
characterized by the trace of the $Q$-matrix,
\begin{equation}
  \label{eq:constraint}
  {\rm   Tr } Q_{\textbf{l}} = 2W_{\textbf{l}}.
\end{equation}

Restricting ourselves to the massless mode
manifold, Eq.~(\ref{eq:parametrization}), we expand the last term in
Eq.~(\ref{eq:F2}) to linear order in
$\hat{\epsilon}$ and $V^0$ and to second order in $\hat{t}$,
\begin{eqnarray}
{\cal F}_d = {\cal F}_0[0] -\frac{g_{T}}{8}
\sum\limits_{\textbf{l}\textbf{l}'}
{\rm Tr} [ e^{i\hat{\chi}_{\bf{ll}'}}
Q_{\bf{l}}e^{-i\hat{\chi}_{\bf{ll}'}}Q_{\bf{l}'}]   \nonumber \\
 +   \frac{\pi}{\delta}
\sum \limits_{j\textbf{l} }\left(\frac{(V^0_{j{\bf l}})^2}{\pi T}
+ {\rm Tr}' [(\hat{\varepsilon}+V^0_{j{\bf l}}
)P_j(Q_\textbf{l}-\Lambda)] \right),
  \label{eq:F3}
\end{eqnarray}
where $\hat{P}_j$ is the projection operator onto the $j$-th replica.
Substituting Eq.~(\ref{eq:F3}) into Eq.~(\ref{eq:Z_abc}), integrating over
$V^0_{j{\bf l}}$ and taking into account that $\textrm{Tr}'[P_j\Lambda]=0$
we obtain the $\sigma$-model for a granular metal,
\begin{subequations}
\label{sigmamodel_ab}
\begin{eqnarray}
\label{eq:sigma_model_Z}
 \langle Z^\alpha \rangle &=& \sum\limits_{\{W\}}
\int
d[Q, \chi]
\exp (-F_d-F_C), \\
\label{eq:F_d}
F_d &=&  -\frac{\pi}{\delta}
\sum\limits_{\textbf{l}}{\rm Tr} [\hat{\varepsilon}Q_\textbf{l}] +
 \frac{\delta }{ 4 T }
 \sum \limits_{j\textbf{l}} N_{j\textbf{l}}^2 \nonumber \\
 && - \frac{g_{T}}{8} \sum\limits_{\textbf{l}\textbf{l}'}
{\rm Tr} [ e^{i\hat{\chi}_{\bf{ll}'}}
Q_{\bf{l}}e^{-i\hat{\chi}_{\bf{ll}'}}Q_{\bf{l}'}], \\
\label{eq:F_C}
F_C&=& \sum\limits_{j\textbf{l}\textbf{l}'}  \frac{ C_{\textbf{l}
  \textbf{l}'}}{2e^2}
\int\limits_0^{\beta}d\tau
\left( \dot{\chi}_{j\textbf{l}} \dot{\chi}_{j\textbf{l}'}-
 \frac{ \delta^2}{4}  N_{j\textbf{l}} N_{j\textbf{l}'}
\right)  .
\end{eqnarray}
\end{subequations}
Here the trace is taken over the
spin, replica, and Matsubara indices.
The summation in Eq.~(\ref{eq:sigma_model_Z}) goes over the integers,
$W_{\textbf{l}}$ in each grain, see Eq.~(\ref{eq:constraint}).
In Eq.~(\ref{sigmamodel_ab}) introduced the notation,
\begin{equation}
\label{eq:N}
N_{j\textbf{l}}=-i\frac{\pi T}{\delta}{\rm Tr}[Q_{\bf{l}}\hat{P}_j]-
q_{{\bf l}}.
\end{equation}
Equation (\ref{eq:N}) has a very natural interpretation: the
$Q$-matrix parametrizes fluctuations of electron Green's function,
and $N_{j\textbf{l}}$ represents the deviation of the charge of
the $\textbf{l}$-th grain in replica $j$ from $q_{{\bf l}}$ due to
$Q$-matrix fluctuations.

We emphasize that we did not have to assume $T \gg \delta$ in
order to integrate out the static component of the auxiliary field
$V^0_{j{\bf l}}$ for a \textit{fixed} value of the $Q$-matrix.
The $\sigma$-model, Eq.~(\ref{sigmamodel_ab}), is valid for any
relation between $T$ and $\delta$. Its action differs from those
in Refs.~\cite{Efetov,Lerner01} by the second term in
Eq.~(\ref{eq:F_d}). This term describes charge neutrality.  It is
instructive to show how this term manifests itself in the
description of charging effects.

First let us consider isolated grains, $g_T = 0$, and set
$q_{j\textbf{l}} \to 0$. In this case the integral over the phases
$\chi$ decouples from the $Q$-matrix integral.  The action in
Eq.~(\ref{eq:F_d}) is minimized when the $Q$-matrix is diagonal in
replica, Matsubara, and spin space, $Q_{\textbf{l}}=\Lambda=
\delta_{ij}\delta_{nm}\delta_{\mu \nu} \textrm{sign}(\epsilon_n)$,
where $\epsilon_n=\pi T (2n+1)$ are fermionic Matsubara
frequencies.  Using the standard parametrization for the
$Q$-matrix~\cite{Efetov97}, $Q = e^{i\hat{u}}\Lambda
e^{-i\hat{u}}$, we observe that the expansion the second term in
Eq.~(\ref{eq:F_d}) starts with the fourth order in $\hat{u}$. At
$T \gg \delta$ the fluctuations of $\hat{u}$ are small, $\langle
\hat{u}^2 \rangle \sim \delta/T$, and the second term in
Eq.~(\ref{eq:F_d}) can be treated perturbatively. As a result we
obtain the following correction to the average free energy,
\begin{equation}
\label{correction}
\Delta F = - \frac{\delta}{4} \left(\frac{\pi T}{ \delta}\right)^2
\lim _{\alpha
\to 0}\frac{1}{\alpha}  \sum\limits_{{\bf l},j}
\left \langle \left({\rm Tr}[Q_{\bf l}\hat{P}_j]\right)^2\right\rangle_0.
\end{equation}
Here $\langle \ldots \rangle_0$ denotes the averaging with respect
to the zero-dimensional action given the first term in
Eq.~(\ref{eq:F_d}). Equation (\ref{correction}) has a clear
meaning.  Coulomb interaction enforces charge neutrality. Thus the
grain charge is independent of the disorder realization is
determined by the external electrostatic potential.  Therefore
ensemble averaged thermodynamic properties can be replaced by
\textit{canonical} averages for non-interacting electrons. It is
well known that for the canonical ensemble mesoscopic fluctuations
result in the following correction to the average free energy of
the grain~\cite{Imry90,Schmid91},
\begin{equation}
  \label{eq:canonical}
  \Delta F = (\delta/4) \langle (\Delta N)^2\rangle_\mu,
\end{equation}
where $\Delta N$ is the fluctuation of the number of particles in the
grain at a constant chemical potential $\mu$. The correction
(\ref{eq:canonical}) is crucial for some quantities, such as persistent
currents~\cite{Ambegaokar90,Imry90,Schmid91,Altshuler91} The quantity $-i
\frac{\pi T}{\delta}{\rm Tr}[Q_{\bf l}\hat{P}_j]$ gives the fluctuation of
the number of particles in the $j$-th replica in grain $\textbf{l}$, see
Eq.~(\ref{eq:N}). Therefore Eq.~(\ref{correction}) is equivalent to
Eq.~(\ref{eq:canonical}). Perturbative evaluation of
Eq.~(\ref{correction}) is straightforward and gives the result of
Ref.~\cite{Schmid91} for a zero-dimensional system, $\Delta F = -\delta
/(2\pi^{2}) \int\int d\epsilon d\epsilon' (\epsilon -\epsilon')^{-2}
f(\epsilon)f(\epsilon')$, with $f(\epsilon)$ being the Fermi function.

Apart from the perturbative correction, Eq.~(\ref{correction}), the
presence of the second term in Eq.~(\ref{eq:F_d}) leads to topological
consequences which are important for the description of Coulomb blockade
effects. In the absence of tunneling the action in Eq.~(\ref{eq:F_d}) has
a discrete set of degenerate minima which are achieved at diagonal
$Q$-matrices, $Q_{\textbf{l}}(\epsilon)=\delta_{ij}\delta_{nm}\delta_{\mu
  \nu}\textrm{sign}(\epsilon_n-2\pi T w_{j\textbf{l}})$, where
$w_{j\textbf{l}}$ are integers analogous to the winding numbers in
the AES approach. It is easy to check the degeneracy of these
minima as the dependencies on $w_{j\textbf{l}}$ in the first and
second terms in Eq.~(\ref{eq:F_d}) exactly cancel each other. This
is a straightforward consequence of gauge invariance since the
different minima are related by singular gauge transformations,
$Q(\tau,\tau') \to e^{i2 \pi Tw \tau
  }Q(\tau,\tau')e^{-i2 \pi Tw \tau' }$.  The sum $2\sum_j
w_{j,\textbf{l}}= W_{\textbf{l}}$, where the factor of two arises from
spin degeneracy, defines the trace of the $Q$-matrix via
Eq.~(\ref{eq:constraint}). Different minima with the same $W_{\textbf{l}}$
belong to the same manifold of $Q$-matrices and can be connected by
continuous rotations in the replica and Matsubara space.

At $T\gg \delta$ the $Q$-matrix fluctuations around the diagonal minima
are suppressed.  One can integrate over them in the Gaussian
approximation, and replace the integral over $Q$ in
Eq.~(\ref{eq:sigma_model_Z}) by the sum over the winding numbers
$w_{j\textbf{l}}$. This summation is equivalent to the summation over all
charge states of the grain~\cite{Shoenreview}.  If $g_T$ is finite but the
temperature is much higher than the elastic escape rate from the grain, $T
\gg \Gamma \sim g_T \delta$, the tunneling term in Eq.~(\ref{eq:F_d}) may
be considered as a perturbation. It merely breaks the degeneracy of the
minima. In this limit one recovers the action of Ref.~\cite{AES} from
Eq.~(\ref{sigmamodel_ab}). Coulomb blockade effects in this regime are
described by discontinuous changes of the winding numbers
$w_{j,\textbf{l}}$ which occur at the tunneling contacts between grains
and correspond to instantons of Ref.~\cite{Korshunov}.

At $T < \Gamma$ the tunneling term in Eq.~(\ref{eq:F_d}) may not
be treated perturbatively.  Since this term is sensitive only to
the spatial variations of the $Q$-matrix and of the phases $\chi$
it preserves the degeneracy of spatially uniform minima of the
action in Eq.~(\ref{eq:F_d}); $\chi =0$ and
$Q=\delta_{ij}\delta_{nm}\textrm{sign}(\epsilon_n-2\pi T
w_{j\textbf{l}})$. The $Q$-matrix may fall into different minima
in different spatial domains.  If the sum $2\sum_j
w_{j\textbf{l}}= W_{\textbf{l}}$ in the two domains is the same
the two minima belong to the same $Q$-matrix manifold. The
$Q$-matrix configuration which minimizes the action,
Eq.~(\ref{eq:F_d}), at the boundary between two such domains
corresponds to a gradual soliton-like rotation of the $Q$-matrix
in replica and frequency space.  The precise form of such solitons
depends on the dimensionality of the system. Their spatial extent
can be estimated from Eq.~(\ref{eq:F_d}) to be the temperature
diffusion length, $L_T \sim a \sqrt{\Gamma/T}$, where $a$ is the
grain size.  At $T \ll \Gamma$ the solitons are delocalized over
many grains and can be described by the continuum limit of the
$\sigma$-model, Eqs.~(\ref{eq:F_d_cont}) and (\ref{eq:F_C_cont}).
Such solitons are analogous to those found in
Ref.~\cite{Kamenev00}.

The $\sigma$-model, Eq.~(\ref{sigmamodel_ab}) has a well defined continuum
limit.  It is obtained by assuming slow spatial variations of the
$Q$-matrix and of $\hat{\chi}=\delta_{ij}\chi_j$, expanding the last term
in Eq.~(\ref{eq:F_d}) to second order in gradients, and replacing
summation over $\textbf{l}$ by integration over $\textbf{r}=a\textbf{l}$.
The tunneling term in Eq.~(\ref{eq:F_d}) in the continuum limit describes
diffusion with the diffusion constant $D=\gamma g_T \delta a^2$, where the
coefficient $\gamma$ depends on the spatial arrangement of the grains,
\begin{eqnarray}
 F_d &=&  \pi \nu \int d\textbf{r}\left[ \sum \limits_{j } \frac{ \mu_{j {\bf r}}^2}{\pi T}  - {\rm
Tr} \left ( \hat{\varepsilon}Q_{{\bf r}} -
\frac{D}{4}[\vec{\partial} Q_{ {\bf r}}]^2 \right)
\label{eq:F_d_cont}
 \right],\\
\label{eq:F_C_cont} F_C&=& \int
\frac{d\textbf{r}d\textbf{r}'}{a^{2d}}
  \int_0^{\beta}d\tau \frac{  C_{\textbf{r}
  \textbf{r}'}}{  2e^2}
\sum\limits_{j}
\left( \dot{\chi}_{j\textbf{r}} \dot{\chi}_{j\textbf{r}'}-
  \mu_{j\textbf{r}} \mu_{j\textbf{r}'}
\right)  ,
\end{eqnarray}
where  $\vec{\partial} Q_{ {\bf r}}= \nabla Q_{ {\bf r}} -i[\nabla \hat{\chi}_{ {\bf r}},Q_{ {\bf r}}]$, and $\mu_{j\textbf{r}}$ denotes the local shift of chemical potential induced
by the fluctuation of the particle number, $N_{j\textbf{r}}$,
\begin{equation}
\label{eq:mu}
\mu_{j\textbf{r}}= \frac{\delta}{2}N_{j\textbf{r}}=-i\frac{\pi T}{2}{\rm
  Tr}[Q_{\bf{r}}\hat{P}_j]- \frac{\delta}{2} q_{{\bf r}}.
\end{equation}
As ${\rm Tr}Q$ is an integer, see Eq.~(\ref{eq:constraint}), it
must remain constant in space in the continuum limit. Different
$W$'s reflect charge quantization in the whole system. Such
effects may be disregarded in the thermodynamic limit, and the
$Q$-integral should be taken over the traceless $Q$-matrices.

In conclusion, we derived the $\sigma$-model for granular metals,
Eq.~(\ref{sigmamodel_ab}).  It describes charging effects in
granular metals in the low temperature regime which includes $T <
\delta$.  The $\sigma$-model action (\ref{sigmamodel_ab})
contains a novel term whose presence is crucial for describing
charging effects. For an isolated grain perturbative treatment of
this term reproduces the canonical ensemble result
(\ref{eq:canonical}).
 In the limit $\delta \to 0$ the $\sigma$-model reproduces the
 action of Ref.~\cite{AES}. In this limit the Coulomb blockade effects are
described by phase instantons of Ref.~\cite{Korshunov}.  At temperatures
below the escape rate from a grain, $\Gamma \sim g_T \delta$, Coulomb
blockade effects within the $\sigma$-model framework are described by
solitons of the $Q$-matrix similar to those studied in
Ref.~\cite{Kamenev00}. The spatial extent of such solitons is given by the
temperature diffusion length, $L_T=a\sqrt{\Gamma/T}$. Inclusion of
such solitons may be important for studying charging effects in granular
systems at $T < \Gamma$.

\begin{acknowledgments}

  We would like thank I.~Aleiner, B.~L.~Altshuler, A.~Kamenev, A.~Lopatin,
  K.~Matveev and J.~S.~Meyer for illuminating discussions, and acknowledge
  the hospitality of KITP, Santa Barbara. This research was
  sponsored by the NSF Grant DMR-9984002, and by the Packard Foundation.

\end{acknowledgments}


\begin{thebibliography}{99}

\bibitem{Altshuler85} B.~L.~Altshuler and A.~G.~Aronov, in
  \textit{Electron-electron interactions in Disordered Systems},
  Eds. A.~L.~Efros and M.~Polk (North Holland, Amsterdam, 1985).


\bibitem{experiment} A.~Gerber, \textit{et. al.} Phys. Rev. Lett. \textbf{
    78}, 4277 (1997).

\bibitem{Efetov}I.~S.~Beloborodov, K.~B.~Efetov, A.~Altland, and
  F.~W.~J.~Hekking, Phys. Rev. B \textbf{63}, 115109 (2001).

\bibitem{Lerner01} I.~V.~Yurkevich and I.~V.~Lerner, Phys. Rev. B \textbf{
    64}, 054515 (2001).

\bibitem{Efetov02} K.~B.~Efetov and A.~Tschersich, Europhys. Lett.,
  \textbf{59}, 114 (2002); Phys. Rev. B \textbf{67}, 174205 (2003).

\bibitem{Lopatin03} I.~S.~Beloborodov, K.~B.~Efetov, A.~V.~Lopatin, and
  V.~M.~Vinokur, cond-mat/0304448 (2003).

\bibitem{we2002} I.~S.~Beloborodov and A.~V.~Andreev Phys. Rev. B
{\textbf 65}, 195311 (2002).

\bibitem{Kamenev03}  A.~Altland, L.~I.~Glazman and A.~Kamenev,
  cond-mat/0305246.

\bibitem{Shoenreview} G.~Sch\"{o}n and A.~D.~Zaikin, Phys. Rep. {\bf 198}, 237
 (1990).

\bibitem{AES} V.~Ambegaokar, U.~Eckern, and G.~Sch\"{o}n, Phys. Rev. Lett.
  \textbf{ 48}, 1745 (1982).


\bibitem{Korshunov} S.~E.~Korshunov, Pis'ma Zh. Eksp. Theor. Fiz.,
  \textbf{45}, 342 (1987) [Sov. Phys. JETP Lett., \textbf{ 45}, 434
  (1987)].


\bibitem{Averin90} D.~V.~Averin and Yu.~V.~Nazarov
Phys. Rev. Lett. \textbf{65}, 2446 (1990).

\bibitem{Matveev95} A.~Furusaki and K.~A.~Matveev
Phys. Rev. Lett. \textbf{75}, 709 (1995);
Phys. Rev. B \textbf{52}, 16676 (1995).


\bibitem{Aleiner97} I.~A.~Aleiner, P.~Brouwer and L.~I.~Glazman, Physics
  Reports, \textbf{ 358}, 309 (2002).


\bibitem{Kamenev00} A.~Kamenev, Phys. Rev. Lett.  \textbf{85}, 4160 (2000).

\bibitem{Finkelstein83} A.~M.~Finkelstein, {\it Electron liquid in
    Disordered Conductors}, edited by I.~M.~Khalatnikov, Soviet Scientific
  Reviews Vol.~\textbf{ 14} (Harwood, London, 1990).

\bibitem{Efetov97} K.~B.~Efetov, {\it Supersymmetry in Disorder and
    Chaos}, Cambridge University Press, New York (1997).

\bibitem{Ambegaokar90} V.~Ambegaokar and U.~Eckern, Phys. Rev. Lett.
  \textbf{ 65}, 381 (1990).

\bibitem{Imry90} Y.~Imry, in "`Quantum Coherence in Mesoscopic Systems"',
  edited by B. Kramer, Proceedings of the NATO Advanced Institute, 1990
  (Plenum, New York).

\bibitem{Schmid91} A.~Schmid, Phys. Rev. Lett. \textbf{ 66}, 80 (1991).

\bibitem{Altshuler91} B.~L.~Altshuler, Y.~Gefen, and Y.~Imry, Phys. Rev.
  Lett. \textbf{ 66}, 88 (1991).



\end{thebibliography}
\end{document}